\documentclass[aps, prl, twocolumn, superscriptaddress]{revtex4}
\usepackage{amsmath} 
\usepackage{amssymb}
\usepackage{amsthm}
\usepackage{xspace}
\usepackage{graphicx}
\usepackage{paralist}

\usepackage{graphics}

\usepackage{mathpazo}
\usepackage{amsfonts}
\usepackage{latexsym}
\usepackage{mathtools}
\usepackage{listings}
\usepackage[usenames,dvipsnames]{xcolor}

\newtheorem{theorem}{Theorem}
\newtheorem{lemma}[theorem]{Lemma} 
\newtheorem{conjecture}[theorem]{Conjecture} 

\newcommand{\tr}{\operatorname{Tr}}
\newcommand{\C}{\mathbb{C}} 
\newcommand{\I}{\mathbb{I}}

\newcommand{\ket}[1]{| #1 \rangle} 
\newcommand{\bra}[1]{\langle #1 |} 
\newcommand{\braket}[2]{\langle #1 | #2 \rangle}

\newcommand{\abs}[1]{\lvert #1 \rvert}

\usepackage{color,graphicx}

\begin{document} 

\title{A note on the inner products of pure states and their antidistinguishability} 
\author{Vincent Russo}
\affiliation{Modellicity Inc., Toronto, Ontario, Canada, \tt{vincent.russo@modellicity.com}}
\affiliation{Unitary Fund}

\author{Jamie Sikora} 
\affiliation{Virginia Polytechnic Institute and State University, Blacksburg, Virginia, USA, \tt{sikora@vt.edu}} 
 
\date{June 16, 2022}

\begin{abstract}   
A set of $d$ quantum states is said to be antidistinguishable if there exists a $d$-outcome POVM that can perfectly identify which state was \emph{not} measured. 
A conjecture by Havl{\'\i}{\v{c}}ek and Barrett states that if a set of $d$ pure states has small pair-wise inner products, then the set must be antidistinguishable. 
In this note we provide a certificate of antidistinguishability via semidefinite programming duality and use it to provide a counterexample to this conjecture when $d=4$. 
\end{abstract} 

\maketitle
                
The distinguishability of a set of quantum states is of central importance in
the study of quantum computing. Indeed, many fundamental problems can be cast
in terms of how well one can infer the identity of which quantum state one
might be holding. 
Formally, suppose we fix a set of quantum states $\{ \rho_1,
\ldots, \rho_n \}$ and we set up a game where Alice selects one of the them,
hands it to Bob, and his task is to determine which state it is. 
To quantify
\emph{how well} Bob can play this game, it often depends on how Alice selects the
state (e.g. randomly, adversarially, etc.). However, if we put the strict
condition on Bob \emph{having to always give the right answer}, then we get the
necessary and sufficient condition that the states must be pair-wise
orthogonal. 
To argue this, we note that by Born's rule, when measuring a quantum state $\rho$ with a POVM $\{ M_1, \ldots, M_n \}$, the probability of the outcome $i$ is given by $\tr(M_i \rho)$. 
Therefore, if the states in the set are pair-wise orthogonal, then it is easy to find a measurement which never fails (simply use a projective measurement which includes the projections onto their supports). 
On the other hand, suppose we have $(M_1, \ldots, M_n)$ being a perfectly distinguishing POVM, i.e., $\tr(M_i \rho_j) = 0$, or equivalently, $M_i \rho_j = 0$, for all $i \neq j$. 
Then, for $i \neq j$,  
$\tr(\rho_i \rho_j) = \tr \left( \rho_i \left( \sum_k M_k \right) \rho_j \right) = 0$, 
thus the states must be pair-wise orthogonal. 
Therefore, in order to certify that a set of states is \emph{not} perfectly distinguishable, it is sufficient to find two non-orthogonal states in the set. 
Certificates are convenient proof tools since
they show the non-existence of something, which can sometimes be a challenging
task. 
We explore (more involved) certificates for a different distinguishing task in this note. 

Suppose we change the game above and instead of tasking Bob to guess which state he is
given, he has to produce a guess for a state he is \emph{not} given. 
For example, if he is given the state $\rho_1$ and he responds ``the state is \emph{not} 
$\rho_2$'', then this would correspond to a correct guess. 
It is worth mentioning that the point here is not trying to ``be wrong" in guessing the state (which might be an interpretation after the previously discussed game), but rather to \emph{exclude} a state which was not given. 
If Bob is able to play this game and win perfectly, we say that the set of states is \emph{antidistinguishable}.  
Mathematically, this requires an antidistinguishing POVM $\{ N_1, \ldots,
N_n \}$ satisfying $\tr(N_i \rho_i) = 0$, for all $i$. 
Again, the interpretation of the outcome $i$ is ``the state is not $\rho_i$'' (which is why we chose the letter $N$ for the notation of such a POVM). 
It is worth noting that we \emph{must} exclude a state which is in the given set; we cannot have an extra measurement operator which outputs ``I do not know''  which is sometimes allowable in state discrimination tasks.  

Finding nontrivial necessary and/or sufficient conditions governing when a set
of quantum states is antidistinguishable or not is tricky. 
This is in stark contrast
to the simple condition of pair-wise orthogonality for the case of perfect
distinguishability. 
Of course, in the case of antidistinguishability,  we can
always exhibit a measurement and check that it satisfies the defining conditions above. 
However, in the 
case of not being antidistinguishable, this is more challenging since this implies the nonexistence of a particular measurement.
We soon discuss how to find such a certificate (which we put to use in a later discussion). 

Antidistinguishability is an interesting property a set of states may have. 
Relaxing the notion of \emph{perfect} antidistinguishability to the task of ``how antidistinguishable are the states?'' was studied in ~\cite{bandyopadhyay2014conclusive} in which they drew connections to the Pusey-Barrett-Rudolph (PBR) theorem~\cite{PBR}. 
In~\cite{havlivcek2020simple}, the work that inspired this note, the authors used this concept to study communication complexity separations. 
Moreover, they posed an \emph{antidistinguishability conjecture} as a means to prove the existence of a two-player communication game
that can be won with $\log(d)$ qubits but would require a one-way communication of $\Omega(d \log d)$ classical bits, thereby providing a (stronger) exponential separation between classical and quantum communication complexities. 
Their antidistinguishability conjecture is as follows. 

\begin{conjecture} 
\label{conj}
\textup{\cite{havlivcek2020simple}} 
If a set of $d$ pure states 
\begin{equation}
\{\ket{\psi_1}, \ldots, \ket{\psi_d} \} \subset \C^d
\end{equation} 
satisfies 
$\abs{\braket{\psi_i}{\psi_j}} \leq (d-2)/(d-1)$ 
for all $i \not= j$, then the set is antidistinguishable. 
\end{conjecture} 

The conjecture holds for the case of $d = 2$ (trivially) and also {$d = 3$~(from the work of \cite{CavesFS02})}, but was previously not
known to be true for $d > 3$. 
Numerical approaches to search for counterexamples for $d \in \{ 4, 5, 6 \}$ in~\cite{havlivcek2020simple} did not produce any. 

In this note, we provide an explicit counterexample to Conjecture~\ref{conj} when $d = 4$. 
We do this by presenting four $4$-dimensional pure states that are deemed to \emph{not} be
antidistinguishable via a particular semidefinite program from~\cite{bandyopadhyay2014conclusive} (which we detail below), and
yet, do have small pair-wise inner products. 
We obtained this counterexample by first randomly generating a set of pure states according to the Haar measure, then determining whether this set is antidistinguishable via semidefinite programming along with a check to determine if their pair-wise inner products satisfy the bound in the conjecture.  
The specific counterexample presented in this note was found after running more than a million random examples. 
Other counterexamples of this dimension were found, but the one presented here has the greatest optimal value of the semidefinite program that was found via our computational search (and thus is the \emph{least} antidistiguishable, in a sense). 

We also provide 
numerical tools that can be used to study different aspects of the
antidistinguishability conjecture for higher dimensions as well as the general
principle of antidistinguishability on its own~\cite{russo2021antidist}.

\section{A certificate of non-antidistinguishability}  
\label{sec:certificate-of-non-antidist}

We first discuss a semidefinite program (SDP) which is mostly identical to the one in~\cite{bandyopadhyay2014conclusive}. 
The only difference is that we do not need to be concerned with any probabilities with which each state is chosen. 

Suppose we fix a set of quantum states $\{ \rho_1, \ldots, \rho_n \}$ and we
consider the following SDP  
\begin{equation} 
\label{primal}
\alpha := \min \left\{ 
    \sum_{i=1}^n \tr(N_i \rho_i) : 
    \sum_{i=1}^n N_i = \I, N_1, \ldots, N_n \succeq 0 
    \right\}. 
\end{equation} 
Note that the optimal value is indeed attained, hence the use of ``min'', since
the feasible region is compact. 
We see that $\alpha \geq 0$ and, moreover, $\alpha = 0$ if and only if the set is antidistinguishable. 
The dual SDP is given by 
\begin{equation} 
\label{eq:antidist_opt}
\beta := \max \left\{ \tr(Y) : Y \preceq \rho_i, \; \forall i \in \{ 1, \ldots, n \} \right\} 
\end{equation}  
where $Y$ is understood to be Hermitian. 
Strong duality was proven in~\cite{bandyopadhyay2014conclusive}, namely that $\alpha = \beta$ 
and that the dual attains its optimal value
(and hence our use of ``max'' above is justified). 
Therefore, we have the following lemma. 

\begin{lemma} 
\label{lem}
A set of states $\{ \rho_1, \ldots, \rho_n \}$ is not antidistinguishable 
if and only if  there exists a Hermitian matrix $Y$ such that $\tr(Y) > 0$ 
and $Y \preceq \rho_i$, for all $i \in \{ 1, \ldots, n \}$.
\end{lemma}

Now it is straightforward to prove a set of states is \emph{not} antidistinguishable, one
must only exhibit a certificate $Y$ satisfying the conditions above. 
Being able to find this certificate is easy in theory, one can solve the dual SDP 
given in Equation~(\ref{eq:antidist_opt}), and for reasonably small examples 
(say, $d$ up to $1000$) this can be done quickly in practice. 
   
\section{Our counterexample (when $d=4$)} 
\label{sec:counterexample}

Define the following four pure states: 
\begin{equation} \label{eq:counterexample_vectors}
\begin{aligned} 
\ket{\psi_1} = \begin{bmatrix} 
     +0.50127198 - 0.037607i \\ 
    -0.00698152 - 0.590973i \\ 
     +0.08186514 - 0.4497548i \\
    -0.01299883 + 0.43458491i 
\end{bmatrix}, \\
\ket{\psi_2} = \begin{bmatrix} 
    -0.07115345 - 0.27080326i \\ 
     +0.82047712 + 0.26320823i \\ 
     +0.22105089 - 0.2091996i \\ 
    -0.23575591 - 0.1758769i       
\end{bmatrix}, \\ 
\ket{\psi_3} = \begin{bmatrix} 
    +0.31360906 + 0.46339313i \\ 
   -0.0465825 - 0.47825017i \\ 
   -0.10470394 - 0.11776404i \\ 
    +0.60231515 + 0.26154959i 
\end{bmatrix}, \\
\ket{\psi_4} = \begin{bmatrix} 
    -0.53532122 - 0.03654632i \\
     +0.40955941 - 0.15150576i \\ 
    -0.05741386 + 0.23873985i \\
    -0.4737113 - 0.48652564i \\ 
\end{bmatrix}.
\end{aligned}
\end{equation}  
We can easily verify that 
\begin{equation} 
\max_{i \neq j} \{ \abs{\braket{\psi_i}{\psi_j}} \} \approx \; 0.64514235 
< \frac{d-2}{d-1} = \frac{2}{3}. 
\end{equation} 
By solving the dual SDP from
Equation~\eqref{eq:antidist_opt} 
with respect to these four pure states, we can ascertain that 
$\{\ket{\psi_1} \bra{\psi_1}, \ket{\psi_2} \bra{\psi_2}, \ket{\psi_3} \bra{\psi_3}, \ket{\psi_4} \bra{\psi_4} \}$ 
is not antidistinguishable. 
We now use its numerically-found optimal solution and Lemma~\ref{lem} 
to provide a certificate of its non-antidistinguishability. 

Define the Hermitian operator $Y$ on the following page 
(see Equation~\eqref{eq:counterexample_feasible_solution}).  
\begin{figure*}[!htpb]
    \begin{equation}\label{eq:counterexample_feasible_solution} 
\footnotesize{Y = 
\begin{psmallmatrix} 
 -0.002352578004032 & -0.006139429568647 + 0.002253370306853i & -0.004431710991485 - 0.000778124769934i  & 0.004045982033136 - 0.002181583048532i \\ 
 -0.006139429568647 - 0.002253370306853i  & 0.003589384258236 & 0.002517710068163 - 0.002392391795840i & -0.009308704240406 - 0.000168259372307i \\
 -0.004431710991485 + 0.000778124769934i  & 0.002517710068163 + 0.002392391795840i & -0.002123263811620 & -0.001232775598439 + 0.000491834467627i \\ 
  0.004045982033136 + 0.002181583048532i & -0.009308704240406 + 0.000168259372307i & -0.001232775598439 - 0.000491834467627i  & 0.001280270586279  
\end{psmallmatrix}}
\end{equation}
\end{figure*} 
Observe that
\begin{equation} \label{dualval}
    \tr(Y) \approx 
0.0003938130288630194 > 0.
\end{equation}
We now wish to show that 
$\ket{\psi_i} \bra{\psi_i} - Y \succeq 0$ 
holds for each $i \in \{ 1, 2, 3, 4 \}$. 
Below we list the eigenvalues of each matrix of interest: 
\begin{equation}
\begin{aligned}
\text{eigs}(\ket{\psi_1} \bra{\psi_1} - Y) = \begin{bmatrix} 
   0.000000000780951 \\
   0.000159290602031 \\
   0.007593054347881 \\
   0.991853848824242
\end{bmatrix}, \\
\text{eigs}(\ket{\psi_2} \bra{\psi_2} - Y) = \begin{bmatrix} 
   0.000000000845682 \\ 
   0.000170622302504 \\ 
   0.006501501274832 \\ 
   0.992934060068367
\end{bmatrix}, \\
\text{eigs}(\ket{\psi_3} \bra{\psi_3} - Y) = \begin{bmatrix} 
   0.000000000751231 \\
   0.000136742588802 \\
   0.009100561906205 \\
   0.990368883698794 \\ 
\end{bmatrix}, \\
\text{eigs}(\ket{\psi_4} \bra{\psi_4} - Y) = \begin{bmatrix} 
   0.000000000905010 \\ 
   0.000186792438756 \\ 
   0.007152857760097 \\ 
   0.992266545011053
\end{bmatrix}. 
\end{aligned}
\end{equation} 
Therefore, $Y$ satisfies all the conditions in Lemma~\ref{lem} implying  the set is not antidistinguishable and thus a counterexample to Conjecture~\ref{conj}.

\section{Supplementary software}  
\label{sec:sosoft}

Supplementary software showcasing the counterexample for $d=4$ may be found at
the following software repository~\cite{russo2021antidist}. The repository
contains Python code that makes use of the Picos Python
package~\cite{sagnol2012picos} to invoke the CVXOPT
solver~\cite{vandenberghe2010cvxopt} for the SDP in 
Equation~\eqref{eq:antidist_opt}.

The set of vectors from Equation~\eqref{eq:counterexample_vectors} were
generated randomly according to the Haar distribution. 
The authors
in~\cite{havlivcek2020simple} followed a similar approach; we simply left our search algorithm running for a very, very long time 
\footnote{Our initial approach to this work was to prove the conjecture was true. 
In the background, we simply left a random search running and running and running  and running and running and running 
to gain intuition from hopefully illustrative numerically-found examples.}. 
The states provided in the counterexample were found after millions of
Haar-random states were generated. Indeed, other such examples were found in
this search as well, but the set of states provided here yielded the highest
value for $\tr(Y)$ (see Equation~\eqref{dualval}). 
The software
from~\cite{russo2021antidist} also allows the user to generate a random
collection of $d$ $d$-dimensional pure states and check whether they are
antidistinguishable by solving the SDP in
Equation~\eqref{eq:antidist_opt}. These numerical tools may be of interest to
further study the notion of antidistinguishability for larger values of $d$. 
On this note, we leave it as an open problem to find the optimal threshold on the 
inner products when $d=4$ and, in general, for larger values of $d$. 
  
\smallskip  
  
\section{Acknowledgements}  

This research was supported in part by the Canadian SR\&ED program. 
VR thanks Vojt{\v{e}}ch Havl{\'\i}{\v{c}}ek and Jonathan Barrett for enlightening
discussions on their paper that inspired this note. 
We also thank 
Srinivasan Arunachalam, 
Basil Singer, 
Ralph Minderhoud, 
and Abel Molina 
for interesting conversations about 
antidistinguishability. 

\bibliographystyle{alpha}
\bibliography{refs}

\begin{thebibliography}{BJOP14}

\bibitem[BJOP14]{bandyopadhyay2014conclusive}
Somshubhro Bandyopadhyay, Rahul Jain, Jonathan Oppenheim, and Christopher
  Perry.
\newblock Conclusive exclusion of quantum states.
\newblock {\em Physical Review A}, 89(2):022336, 2014.

\bibitem[CFS02]{CavesFS02}
Carlton~M. Caves, Christopher~A. Fuchs, and R\"udiger Schack.
\newblock Conditions for compatibility of quantum-state assignments.
\newblock {\em Phys. Rev. A}, 66:062111, 2002.

\bibitem[HB20]{havlivcek2020simple}
Vojt{\v{e}}ch Havl{\'\i}{\v{c}}ek and Jonathan Barrett.
\newblock Simple communication complexity separation from quantum state
  antidistinguishability.
\newblock {\em Physical Review Research}, 2(1):013326, 2020.

\bibitem[PBR12]{PBR}
Matthew~F. Pusey, Jonathan Barrett, and Terry Rudolph.
\newblock On the reality of the quantum state.
\newblock {\em Nature Physics}, 8(6):475--478, 2012.

\bibitem[Rus21]{russo2021antidist}
Vincent Russo.
\newblock antidist: A {P}ython toolkit for studying the antidistinguishability
  conjecture.
\newblock \url{https://github.com/vprusso/antidist}, November 2021.

\bibitem[SS12]{sagnol2012picos}
Guillaume Sagnol and Maximilian Stahlberg.
\newblock Picos, a {P}ython interface to conic optimization solvers.
\newblock In {\em Proceedings of the in 21st International Symposium on
  Mathematical Programming}, 2012.

\bibitem[Van10]{vandenberghe2010cvxopt}
Lieven Vandenberghe.
\newblock The {CVXOPT} linear and quadratic cone program solvers.
\newblock {\em Online: http://cvxopt. org/documentation/coneprog. pdf}, 2010.

\end{thebibliography}
  
\end{document}